# A NOVEL REAL-TIME GEOLOCATION TRACKING TOOL

**Erkan Meral[1], Mehmet Serdar Güzel[2]**

[1] *Computer Engineering Dept. of Ankara University, Ankara, TR*
[2] *Computer Engineering Dept. of Ankara University, Ankara, TR*
*Corresponding author: mguzel@ankara.edu.tr*

## Abstract

Global Positioning System (GPS) is a satellite network that transmits regularly encoded information and makes it possible to pinpoint the exact location on Earth by measuring the distance between satellites and the receiver. While GPS satellites continually emit radio signals, receivers are able to receive these signals. This study proposes a tool in which an electronic circuit that is consisted of integration of SIM908 shield and Arduino card is used as a GPS receiver. The positional data obtained from GPS satellites yields error due to the noise of the signals. Accordingly, in this study Kalman and Average filters are applied respectively in order to reduce these faults and handle the overall positional error. Several experiments were carried out in order to verify the performance of the filters within the GPS data. The results of these enhanced systems are compared with the initial configuration of the system severally. Especially the results obtained using the Kalman filter is quite encouraging.

## 1. INTRODUCTION

Estimation of exact object positions located on the Earth is a crucial problem GPS is usually used for such positioning operations. A unique system based on GPS technology and Arduino card were introduced at authors' previous paper [1]. The flowchart of the proposed system is illustrated in Figure 1. This study proposes a system that processes data obtained from GPS sensor integrated into an electronic circuit, which in essence, estimates real time locations and illustrates them on a map via a web site. However, the accuracy of position data obtained from GPS includes some error margin. This error margin can easily be increased once the GPS signals are obtained in a noisy environment. These environments may cause distortions, mainly occurring in signal lines, must be prevented in order to increase overall reliability of the data. One of the most efficient way of noise removing is to apply proper filtering techniques. One of the most suitable filters for signal enhancement is the Kalman filter [2]. The goal of implementing the Kalman filter is to prevent large changes in the signal from suddenly appearing in the signal from affecting the system at a large scale, starting with the previous data obtained from the system. For this study, an appropriate Kalman filter is applied to data so as to prevent sudden changes in the positions obtained by the proposed device [3]. The main reason lies behind this post-processing operation is to prevent these undesirable changes in the system that allows to obtain more accurate and practical resultant position data. Alternatively, one of the filters that is applied in many operations where the signal processing operation is carried out is the Average filter [4].

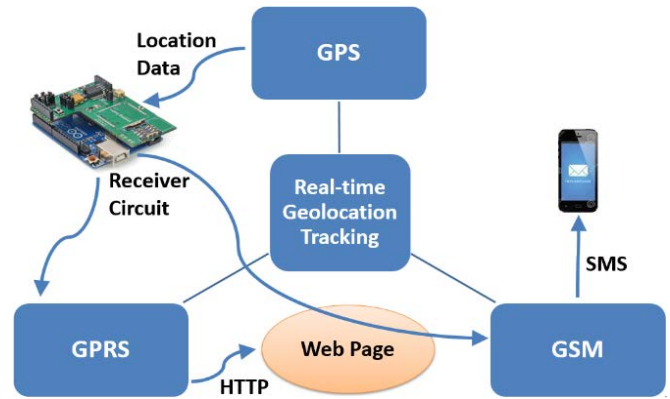

**Figure 1**. Architecture of the proposed real-time geolocation tracking and geofencing system [1].

This filter is normally used to eliminate sharp transitions on images. As well as, it is mainly designed based on the arithmetic average calculation on the pixel values in the image, and since it is a general filter, it can be employed for different problems [5]. Accordingly, in this study an appropriate average filter is also applied to improve GPS positions obtained from the proposed device. Overall, this study incorporates two different filtering approaches into the proposed device separately in order to enhance position data and increase complete accuracy of the proposed device. While Section 2 details proposed filtering techniques, Section 3 illustrates experimental results and finally, the paper is concluded in Section 4.

## 2. Methodology

This section proposes two different post-processing algorithm based on Kalman and Average filtering respectively. These filtering approaches and also example scenarios for each approach are detailed in the following paragraphs.

## 2.1 Kalman Filter for GPS data enfacement

Kalman filter is reliable and efficient way of noise removal process for especially signal and image processing fields. The position data obtained by the proposed device consists of the latitude and longitude data [2]. Accordingly, faults occurring in the location data are the result of faults occurring in the values of latitude and longitude data. Hence, the Kalman filter is applied separately on the latitude and longitude values. The new latitude and longitude values, enhanced by the help of filtering process, are reassembled to determine the new location, which is far more accurate than the previous measurements as it is expected. The Kalman filter offers complicated equations for different problems and systems. When the filter is applied, the appropriate equations for each problem should be taken into account so as to reduce overall complexity of the system. Essentially, the matrices that are not needed in equations can be omitted. In this study, the following formula (1) is obtained by subtracting the unused condition matrices from the main formulas when the Kalman filter is required to apply for the signal processing problem

$$X_k = X_k . Z_k + (1 - K_k) . X_{k-1} \quad (1)$$

According to the previously given equation, k, is used as a sub-index in the form, denoting the operating states of the system. The purpose of the formula is to compute the predicted $X_k$ values of the signal. $Z_k$ value is the original value obtained from the receiver continuously which encompasses a certain amount of error. Besides, $X_{k-1}$ denotes the estimated value of the signal of the previous state, whereas The $K_k$ value is called the Kalman gain. This is the only unknown value in the equation and for each case it is recalculated with the following formula (2) based on the values of the previous error covariance $P_k$ and the standard deviation of the measurement R values.

$$K_k = P_k / (P_k + R) \quad (2)$$

In cases where the Kalman gain is taken as a constant value of 0.5, the equation will behave as if it is an average filter [6]. By recalculating the Kalman gain in each step, the optimum average value can be calculated and also the capabilities of the Kalman filter will be used [7]. In order to calculate the previously mentioned values preferred in this algorithm, some initial values and parameters need to be determined. The standard deviation of measurement R value to be used as recalculating the Kalman gain at each step is set to 1 in practice. For $X_k$, the first estimated value $X_0$ to be used at time k=0 is set as the first position data received from the device. That is, when the Kalman filter is applied on latitude values, while the first position received from the device is assigned to the latitude value $X_0$, while the filter is applied on the longitude values, the longitude value of the first position received from the device is assigned to the value $X_0$. The first value to be used at time k=0 of the $P_k$ value used as error covariance $P_0$ value is set to 4. This value can be set to a non-zero value. Setting $P_0$ to zero means that there is no noise in the environment. The $P_k$ value for each case will be recalculated using the Kalman gain value, $K_k$, and the previous error covariance value $P_k$ using the following formula (3).

$$P_k = (1 - K_k) . P_k \quad (3)$$

Where, $P_k$ refers the Kalman gain, $K_k$ denotes Kaltman gain. Essentially the Kalman filter is executed within the following algorithm.

*Algorithm 1: GPS enhancement problem using 1-D Kalman Filter*

---

*for k = 0 to 30*
 $Z_k$ ← *Receiver_Values[k]*
 $X_k$' ← $X_k$
 $P_k$' ← $P_k$
  $K_k$ ← $P_k$' / ($P_k$' + R)
  $X_k$ ← $K_k$ * $Z_k$ + (1 - $K_k$) * $X_k$
  $P_k$ ← (1 - $K_k$) * $P_k$'
 *Kalman_Values[k]* ← $X_k$
*end-for*

---

The steps of applying the Kalman filter to an example latitude values from obtained from the receiver device is illustrated on the Table 1, at times k=0 and k=1. The position data obtained from the receiver consists of the latitude and longitude data as previously mentioned. In order to defeat faults the Average filter can also be applied separately to latitude and longitude values. The new latitude and longitude values obtained from this filter based enhacement addresses more accurate positon data. The Average filter applied for the given problem is defined as follows. First, 30 latitude data of 30 position data obtained from the GPS receiver are recorded in an array. The result of applying the Average filter to this latitude data will result in another array of 30 elements. The first element in the array in which the data from the receiver is held will be the same as the first element in the array in which the Average filter results are held. The second element in the array where the filter results are held will be the arithmetic average of the first two elements in the array in which the receiver records are held. In the same way, the third element in the array where the filter results are held will be the arithmetic mean of the first three elements in the array in which the receiver records are held. In other words, it will be progressed gradually by averaging over cumulative totals. The Average filter is run with the following algorithm.

*Algorithm 2: GPS enhancement problem using Average Filtering*

---

*for k = 0 to 30*
  *Total* ← *Total + Receiver_Values[k]*
  *Value* ← *Total / (k + 1)*
  *Average_Values[k]* ← *Value*
*end-for*

---

The above loop runs 30 times. At the end of the loop, the latitude and longitude values from the gathered from the receiver are separately filtered by the Average filter. Afterwards, all new latitude and longitude values obtained from the filter result are recorded in to the Average Values array. Table 2 illustrates a numeric example using this filtering technique based on real measurement values.

**Table-1:** An example scenario using Scenario 1 based on Kalman Filter

| | | |
|---|---|---|
| $R$ | 1 | 1 |
| $k$ | 0 | 1 |
| $Z_k$ | 39,953250 | 39,953200 |
| $X_k$ | 39,953250 | 39,953250 |
| $P_k$ | 4 | 0,8 |
| $X_{k-1}$ | 39,953250 | 39,953250 |
| $P_{k-1}$ | 4 | 0,8 |
| $K_k$ | $K_k = P_{k-1} / (P_{k-1} + R)$, $K_0 = 4 / (4 + 1)$<br><br>$K_0 = 0,8$ | $K_1 = 0,8 / (0,8 + 1)$, $K_1 = 0,44$ |
| $X_k$ (new) | $X_k = K_k.Z_k + (1 - K_k).X_{k-1}$<br><br>$X_0 = 0,8.39,953250 + (1 - 0,8).39,953250$<br><br>$X_0 = 39,953250$ | $X_1 = 0,44.39,953200 + (1 - 0,44).39,953250$<br><br>$X_1 = 39,953228$ |
| $P_k$ (new) | $P_k = (1 - K_k). P_{k-1}$<br><br>$P_0 = (1 - 0,8).4$<br><br>$P_0 = 0,8$ | $P_1 = (1 - 0,44).0,8$<br><br>$P_1 = 0,448$ |

**Table-2:** An example scenario using Scenario 1 based on Average Filter

| | | |
|---|---|---|
| $k$ | 0 | 1 |
| Total | 0 | 39,953250 |
| Receiver_Values[k] | Receiver_Values[0] = 39,953250 | Receiver_Values[1] = 39,953200 |
| Total (new) | Total = Total + Receiver_Values [0]<br><br>Total = 0 + 39,953250<br><br>Total = 39,953250 | Total = 39,953250 + 39,953200<br><br>Total = 79,906450 |
| Value | Value = Total / (k + 1)<br><br>Value = 39,953250 / (0 + 1)<br><br>Value = 39,953250 | Value = 79,906450 / (1 + 1)<br><br>Value = 39,953225 |
| Average Values[k] | Average Values[k] = Value<br><br>Average Values[0] = 39,953250 | Average Values[1] = 39,953225 |

# 3. Results and Analysis

Several experiments have been carried out to calculate the position error margin determined by the receiver. It has been determined that the accuracy of the position data given by the receiver during operation is variable during clear and cloudy weather. For this reason, experiments were carried out separately for clear and cloudy weather. However, only clear weather results are shown in this paper due to the scope of the study. For a comprehensive outdoor experiment, illustrated in table 3, the position data were taken 30 times in a clear day. The corresponding table illustrates this scenario that include positon data, satellite numbers and also error margin in meters. The Record Id, Latitude and Longitude columns shown in Table 3, indicating the data obtained from the receiver tool. The Satellites column indicates how many GPS satellites you communicate with when the receiver receives the data. Error Margin column indicates the value of meter in terms of the error rate of the position data obtained by the receiver. To be able to determine the location, the receiver must exchange signals with at least three GPS satellites. However, it should be noted that the number of satellites in the charts is always greater than three. Considering that the sequence is sorted according to the increasing time, it is observed that the amount of the satellite communicating with the receiver increases as time progresses. So the receiver communicates with more satellites over time. Figure 2 illustrates 30 data location obtained from the receiver tool and shown on Google Maps in a clear weather day. The location illustrated with the green colour is the actual location including the receiver location, whereas the locations shown in red are the positions taken for the experiments. According to the Table 3 and Figure 2 it can be easily assumed that as the number of satellites incorporated into the system increases, the error margin decreases gradually. Figures 3 and 4 show the variation of the error margin calculated according to the original data obtained from the receiver using Kalman and Average Filter severally. Once the Figure 3 is evaluated, the error margin in the last recording on the receiver clear weather record is measured as 9,39 meters. When the values are passed through the Kalman filter, the error margin in the same register is measured as 3,64 meters. However, as it is evaluated according to the last records, it can be said that the Kalman filter provides the benefit of 5,65 meters in the clear weather. It is also possible to compare the minimum of the error margin values obtained from the device with the minimum values of the error margin values obtained from the Kalman filter. Minimum error value received from the receiver is 9,39 meters taken in the last record. Minimum error value received from the Kalman is 3,47 meters taken in the $21_{th}$ record. Based on these values, it can be said that the Kalman filter provides a benefit of 5,92 meters to the system. On the other hand, according to the Figure 4, the error margin in the last recording on the receiver clear weather record is measured as 9,39 meters. When the values are enhanced via an Average filter, the error margin in the same register is measured as 4,18 meters. it can be said that the Average filter provides the benefit of 5,21 meters in the clear weather as the last record are considered. Table 4 illustrates error margin comparison between the tool and filters applied in different environmental conditions.

## 4. CONCLUSIONS

This paper introduces post-processing techniques to enhance GPS data obtained from a variety of satellites. Position data is obtained from a device which was previously designed by authors. The device is able to connect satellites and obtain location data using Latitude and Longitude values. However, due to unexpected noise of the signals, there may occur big position errors in terms of approaching the exact location data. In order to reduce this critical error, the corresponding noise must be reduced into a tolerable level. Accordingly, Kalman and Average filters are compared to enhance the overall position estimation performance. According to the results, both filtering approaches have improved overall performance of the systems successfully. However, Kalman Filter has superiority over the Average filter on both clear and cloudy weathers as it is expected.

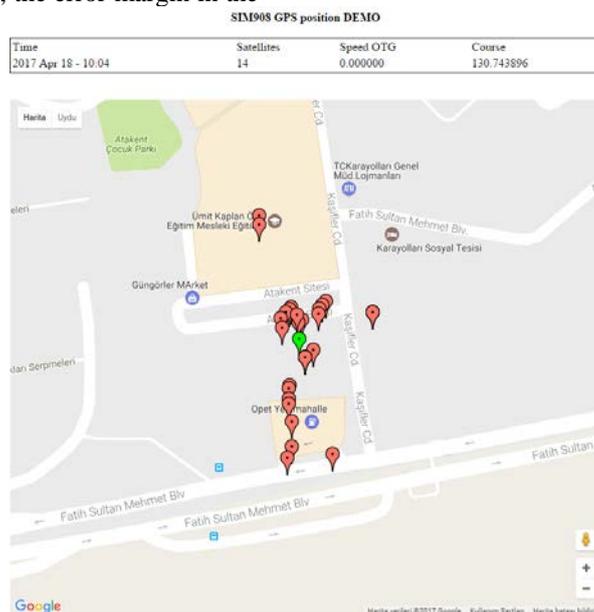

**Figure 2:** 30 data location obtained from the receiver tool and shown on Google Maps in a clear weather.

**Table-3 :** Position values received from the receiver in clear weather Latitude: 39,9525646 - Longitude: 32,7966589

| Record Id | Latitude  | Longitude | Satellites | Error Margin (meter) |
|-----------|-----------|-----------|------------|----------------------|
| 0         | 39,953250 | 32,796365 | 3          | 80,22                |
| 1         | 39,953200 | 32,796365 | 3          | 74,96                |
| 2         | 39,951920 | 32,796900 | 3          | 74,56                |
| 3         | 39,951901 | 32,796570 | 4          | 74,17                |
| 4         | 39,951961 | 32,796604 | 4          | 67,28                |
| 5         | 39,952102 | 32,796604 | 5          | 54,70                |
| 6         | 39,952711 | 32,797189 | 5          | 48,02                |
| 7         | 39,952711 | 32,797189 | 5          | 48,02                |
| 8         | 39,952711 | 32,797189 | 6          | 48,02                |
| 9         | 39,952201 | 32,796580 | 6          | 40,98                |
| 10        | 39,952231 | 32,796580 | 7          | 37,69                |
| 11        | 39,952287 | 32,796580 | 7          | 31,59                |
| 12        | 39,952303 | 32,796589 | 7          | 29,69                |
| 13        | 39,952771 | 32,796851 | 8          | 28,19                |
| 14        | 39,952753 | 32,796824 | 8          | 25,23                |
| 15        | 39,952732 | 32,796801 | 9          | 22,20                |
| 16        | 39,952703 | 32,796797 | 9          | 22,20                |
| 17        | 39,952741 | 32,796598 | 10         | 20,29                |
| 18        | 39,952726 | 32,796588 | 10         | 18,93                |
| 19        | 39,952709 | 32,796561 | 11         | 18,09                |
| 20        | 39,952678 | 32,796525 | 11         | 17,00                |
| 21        | 39,952624 | 32,796534 | 12         | 12,52                |
| 22        | 39,952460 | 32,796701 | 12         | 12,17                |
| 23        | 39,952460 | 32,796701 | 12         | 12,17                |
| 24        | 39,952460 | 32,796701 | 12         | 12,17                |
| 25        | 39,952460 | 32,796701 | 13         | 12,17                |
| 26        | 39,952670 | 32,796678 | 13         | 11,83                |
| 27        | 39,952501 | 32,796761 | 13         | 11,21                |
| 28        | 39,952651 | 32,796653 | 14         | 9,62                 |
| 29        | 39,952648 | 32,796641 | 14         | 9,39                 |

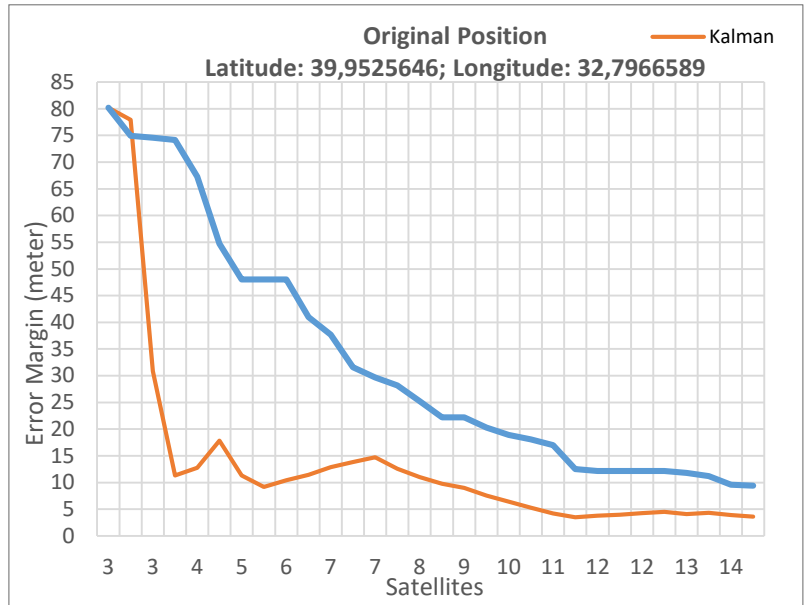

**Figure 3:** Error margin graph of the receiver values and the Kalman filter results in clear weather.

**Table-4:** Error margin comparison between the tool and filters applied in different environmental conditions.

|  | The Receiver Values (meter) | The Kalman Filter Results (meter) | The Kalman Filter Improvement Rate (%) | The Average Filter Results (meter) | The Average Filter Improvement Rate (%) |
|---|---|---|---|---|---|
| **Minimum Error Margin in Clear Weather** | 9,39 | 3,47 | 63,04 | 4,18 | 55,48 |
| **Minimum Error Margin in Cloudy Weather** | 19,50 | 11,76 | 39,69 | 12,29 | 36,97 |

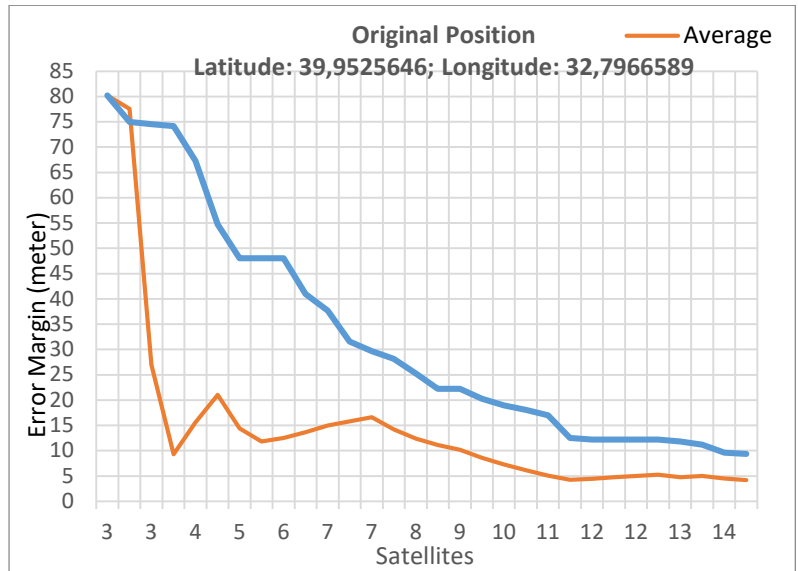

**Figure 4:** Error margin graph of the receiver values and the Average **filter** results in clear weather.